\documentclass[preprint]{aastex}
\usepackage{amsmath,amssymb}

\newcommand{\kms}{\ensuremath{\,\mbox{km}\,\mbox{s}^{-1}}}

\newcommand{\oi}{O\,{\sc i}}

\newcommand{\oiii}{O\,{\sc iii}}

\newcommand{\neiii}{Ne\,{\sc iii}}
\newcommand{\nev}{Ne\,{\sc v}}

\newcommand{\sii}{S\,{\sc ii}}

\newcommand{\fevii}{Fe\,{\sc vii}}

\begin{document}

\title{Are the Narrow Line Regions in Active Galaxies Dusty and Radiation
Pressure Dominated?}

\author{Michael A. Dopita, Brent A. Groves, \& Ralph S. Sutherland}
\affil{Research School of Astronomy \& Astrophysics,
Institute of Advanced Studies, The Australian National University,
Cotter Road, Weston Creek, ACT 2611
Australia}
\email{Michael.Dopita@anu.edu.au, bgroves@mso.anu.edu.au, ralph@mso.anu.edu.au}

\author{Luc Binette}
\affil{Instituto de Astronoma, UNAM, Ap. 70-264, 04510 D.F., Mexico}

\author{Gerald Cecil}
\affil{Dept. of Physics \& Astronomy, U. of N. Carolina, Chapel Hill,
NC 27599-3255}

\singlespace

\begin{abstract}
The remarkable similarity between emission spectra of narrow line regions
(NLR) in Seyfert Galaxies has long presented a mystery.  In photoionization
models, this similarity implies that the ionization parameter is nearly always
the same, about $U \sim 0.01$.  Here we present dusty, radiation-pressure
dominated photoionization models that can provide natural physical insight
into this problem. In these models, dust and the radiation pressure acting on it
provide the controlling factor in moderating the density, excitation 
and surface
brightness of photoionized NLR structures. Additionally, photoelectric heating by the dust
is important in determining the temperature structure of the models.
These models can also explain the coexistence of the low-, intermediate- and
coronal ionization zones within a single self-consistent physical structure.
The radiation pressure acting on dust may also be capable of driving the
fast ($\sim 3000 \kms$) outflows such as are seen in the HST observations
of NGC~1068.
\end{abstract}

\keywords{galaxies: active --- galaxies: Seyfert --- ISM: general ---
line: formation}

\section{\label{Intro}Introduction}

Optical spectroscopy of the ionized gas around galactic nuclei provides
strong constraints on the excitation mechanisms. Line ratio diagrams of
active galactic nuclei (AGN), such as those of \citet{VO87}, allow us to
distinguish three major classes; nuclei excited by starbursts, and the two
types excited by a bona-fide active nucleus: the Seyfert narrow line regions
(NLRs) and extended narrow line regions (ENLRs), and the Low Excitation
Nuclear Emission Line Regions (LINERs). As far as the Seyfert galaxies are
concerned, the compilations of \citet{VO87}, Veilleux(1991a,b,c), and %
\citet{VV00} show that the emission line ratios are remarkably uniform
within a given object, and very similar (with typically less than 0.5 dex
variation) from one object to another.

The standard paradigm proposes that the NLR and the ENLR are excited by
photons originating at or near a compact nuclear source (see, e.g. %
\citet{Ost89}) having a smooth featureless power-law, or broken power-law
EUV ionizing spectrum. Within this model, the clustering of the observed
line ratios within such a restricted domain of parameter space presents a
problem, since in such models the excitation is in large measure controlled
by the ionization parameter, $U$, and this is free to vary over a wide
range. The implications of this is that there should exist photoionized regions
with very high excitation corresponding to high $U$, but objects such as these
are missing in the \citet{VO87} diagnostic diagrams. Instead, the constraints derived from the line ratio diagnostic
diagrams imply that $U$ has a similar value in all NLRs. Modelers have been
therefore forced to make the arbitrary (and possibly unphysical) assumption
that the gas density in the ionized clouds must fall exactly as the inverse
square of the distance from the nucleus.

One rather successful variation on the photoionization models is to assume
that the power-law EUV spectrum passes through a highly-ionized screen of
diffuse gas, before being absorbed in the vicinity of dense (much higher
pressure) clouds (\citet{bws96}, \citet{bwrs97}). Although such models
produce generally good results when compared with observed ratios, these
have the difficulty that the geometry of the ionized gas is ill-defined and
necessarily becomes a new free parameter. These models can also be
questioned on physical grounds, since the assumed pressure difference
between the cloud and inter-cloud components is not explained.

Another variation is to consider the integrated emission from a whole family
of photoionized clouds of different densities (\citet{KomSch97,Ferg97}).
These models can be made to fit the observed spectra quite well, but the
geometry of the clouds and their relative contributions to the emission
remain free parameters and are poorly constrained.

In any of these photoionized models, the (often quite violent) internal
dynamics of the ENLR are not addressed. These include clear evidence of
non-gravitational motion, and evidence for outflow at velocities up to (or
even in excess of 1000 km s$^{-1}$) ({\em eg} \citet{Pedlar89}, %
\citet{Allen99}). It was results such as this that led \citet{Wilson80} to
suggest that the nucleus ejects radio components that interact with ambient
gas and replenish the high kinetic energy and ionization of the NLR. This
led to the development of photoionization by fast radiative shocks, which
generate a strong internal photoionizing radiation field (\citet{do:su95},
(1996)). This model is successful in reproducing the strong lines observed
in many AGN (see review by \citet{Do00}). In addition, it can explain other
features which simple photoionization models fail to address. For example,
where the NLRs are spatially resolved, there often are found to be strong
correlations between radio power and either line luminosity \citet{deBruyn78}
or line width \citet{Wilson80}. Such correlations exist not only for Seyfert
galaxies, but persist up to much more luminous classes of radio galaxies.
However, the fast shock model has its limitations. It has difficulty
reproducing the very strong coronal lines of [Fe\thinspace {\sc vii}],
[Fe\thinspace {\sc x}], [S\thinspace {\sc viii}], [S\thinspace {\sc xii}],
[Si\thinspace {\sc vi}] or [Si\thinspace {\sc vii}] (\citet{wilray99}) which
imply photoionization with very high local ionization parameters. These
models also cannot reproduce the large outflow velocities inferred in the
[O\thinspace {\sc iii}] line in some objects such as NGC\thinspace 1068
(e.g. \citet{KC2000,Cecil01}). In shock models, most of the [O\thinspace
{\sc iii}] emission is expected to be generated in the photoionized
precursor of the shock, which theoretically should be the undisturbed gas in
the galaxy. Furthermore, the NLR lines of (probably) the majority of Seyfert
galaxies are quite narrow and consistent with line broadening due to
galactic rotation rather than being produced in shocked outflows.

In this paper we argue that some of these properties can be explained in
terms of radiation pressure dominated dusty photo-ablating clouds. In the
past, dust has been considered in photoionization models both through its
contribution via photoelectric heating (e.g.~\citet{Bott98}) and as
shielding component thanks to its large opacity in the FUV. However, the
effects of dust photoelectric heating and of the radiation pressure on dust
on both the ionization and hydrostatic structure of the NLR clouds has not
been examined in a systematic fashion before. In the next section we detail
these effects and show how these provide a natural explanation as to why the
inferred NLR ionization parameter seems to be confined to the range $%
-2.5<\log U<-2$. In \S 3 we describe the basics of our dusty, radiation
pressure dominated models. These are compared with other types of
photoionization models, allowing us to separately understand the effects
of photoelectric heating and of radiation pressure on the emergent spectrum.
The predictions of all these models are compared with observations of NLR of
Seyferts on some key emission line diagnostic diagrams. The concluding
remarks are given in \S 4. In following papers we will investigate further the
effects of different parameters on our radiation pressure dominated
photoionization models and present a full hydrodynamical model exploring such
structures. 

\section{\label{paradigm}A New Paradigm: Dusty, Radiation-pressure Dominated
Photoionization}

The effects of radiation pressure from line absorption have previously been
considered in photoionization models of the emission line regions of AGN (%
\citet{EliFer86}). This is important in determining the internal density
structure at large ionization parameter. The effect of dust within the ENLR
has also been considered previously (\citet{NL93}). However, the radiation
pressure exerted by the ionizing source acting upon dust within the
photoionized gas has usually not been accounted for by the standard
photoionization models, with a few notable exceptions (e.g.~\citet{Baldwin91}%
). The standard photoionization modelling assumes an isochoric (or constant density)
structure within the ionized slab instead of an isobaric (constant pressure)
structure. Dust is important not only because of the 
radiation pressure acting upon it, but also because of the photoelectric
heating which it produces in the photoionized plasma, and in the way it
competes with the gas in absorbing the ionizing photons. 

However, all of
this critically depends on whether dust can survive the either the strong
EUV radiation field, or the relatively energetic conditions characterizing
the photoionized region. In the ionized plasma of the ENLR, grains are
electrically charged, locking the motion of the dust to that of the
ionized gas. Both the ionized gas and the dust is likely to be streaming at
roughly sonic velocities from ionization fronts located around
photo-evaporating dense clouds located in the ``ionization cones'' of the
AGN. The problem of dust survival is therefore reduced to the question of
whether it can survive for a time equal to its passage through the
high-emissivity region near the cloud surface as discussed in \S 3. Our
detailed photoionization 
models show that the average gas temperature is $\sim 10^4$K, at which the
thermal sputtering of grains is negligible and  incapable of destroying the
grains in so short a timescale. This is in contrast with the case of radiative
shocks in which gas
temperatures reach $\sim 10^5 - 10^6$K and grain destruction by
sputtering occurs on very short timescales (\citet{Tieletal94}). Similarly, our
models show that grain temperatures are too low for thermal evaporation of
grains by quantum heating provided that the dust grains are large enough
($\gtrsim 3$ nm ). 
A third potential dust
destruction mechanism is that of Coulomb explosion, in which the grain
charging becomes large enough that the Coulomb repulsion between the atoms
composing the grain exceeds the atomic bond strength. According to our
models this condition is met for grains smaller than about 10 nm when the
ionization parameter is of order unity. In the models described here we have
simply set the minimum grain radius equal to 10 nm to allow for the
destruction of the smaller grains. We will return to the detailed physics of
grain destruction processes in these environments
in a future paper. 

\begin{figure}
\scalebox{0.8}[0.8]{\plotone{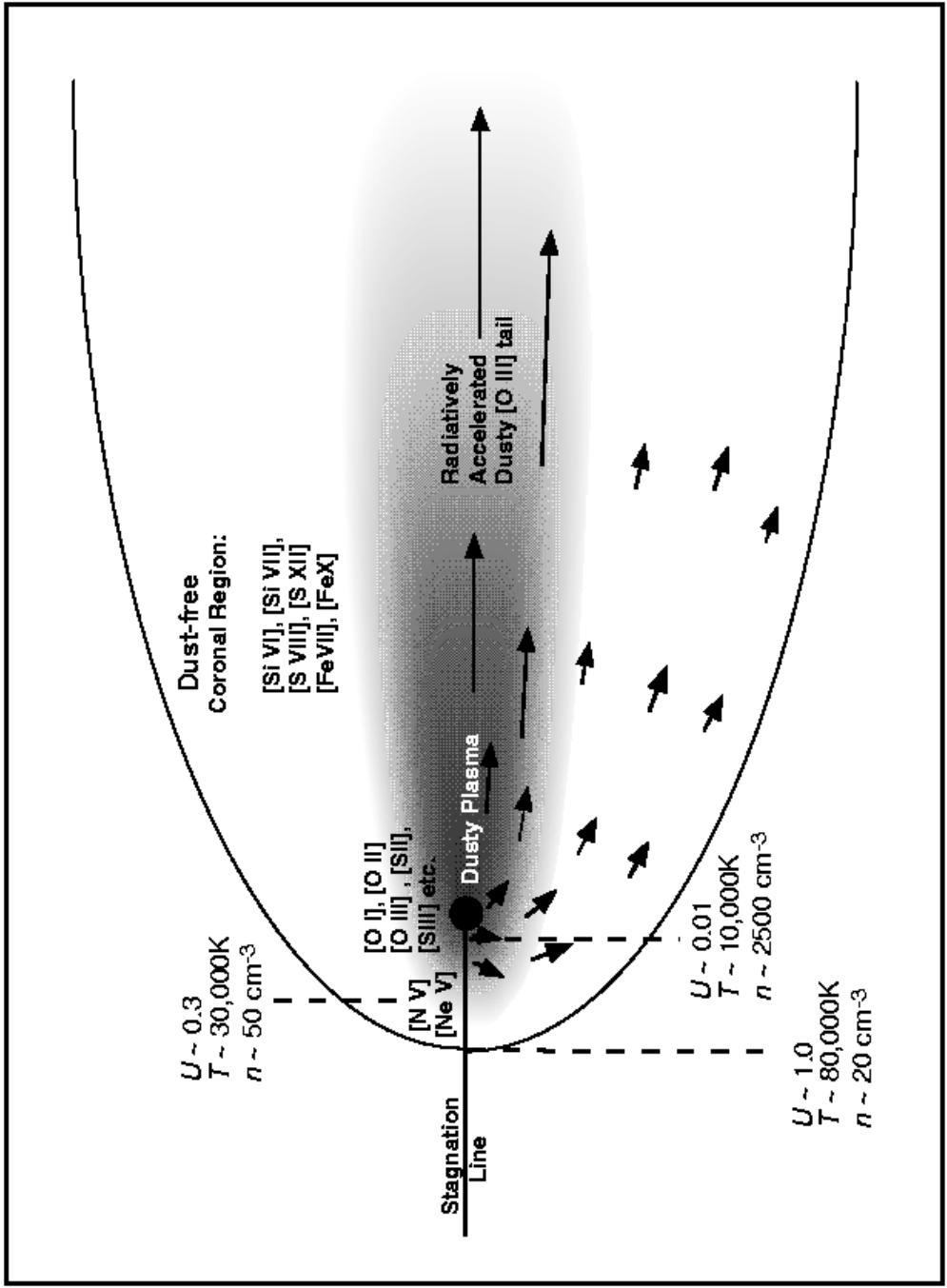}}
\caption{\label{flow} Depiction of our theoretical model showing the
photoevaporating, dusty molecular cloud (black circle) at the center
surrounded by the dusty, 
photoevaporated plasma, from which both the low- (close to the cloud) and
intermediate-ionization (close to the coronal region) lines are emitted. This
plasma is itself surrounded by the dust-free coronal 
region. The stagnation line indicates the region modelled by this paper. Each
region (excluding the molecular cloud) is marked with some of the strong lines
emitted from this part of the 
structure. The parameters marked by the dashed lines indicate typical values
for each region for a NLR cloud found in NGC 1068.}
\end{figure}

In NLR regions observed by \citet{oliv94}, \citet{moor96}, and \citet{KC2000}
the coronal lines appear remarkably strong as compared to H$\beta $. The
existence of such coronal gas requires that both the ionization parameter
and the radiation pressure are very large (\citet{bwrs97}). Elements such as
Fe and Si are normally largely sequestered in the grains. The strength of
the coronal lines of these elements requires that in the coronal gas, at
least, the dust has been largely destroyed. Our model, as depicted in Figure
\ref{flow} is therefore one of a
dusty, radiation-pressure dominated region surrounding a photoevaporating
molecular cloud, which in turn is surrounded by a coronal halo within which the dust
has been largely destroyed. In this paper, we will consider only the
emission spectrum generated by the dusty dense inner region near the
stagnation point in the flow around the cloud. 
This will be the highest
emissivity region in the flow, and it will be the region in which the
radiation pressure gradient is matched by the gas pressure gradient,
allowing us to construct static rather than dynamic photoionization models.

Consider $\Xi _{0}$, the ratio of the available pressure of ionizing
radiation, $\Phi$, to the gas pressure at the irradiated photoionized
surface of a photoevaporating cloud\footnote{%
This definition implies a factor 2.3 lower value than that of \citet{kro:a}}
:
\begin{equation}
\Xi _{0}=\frac{P_{{\rm rad}}}{P_{{\rm gas}}^{0}}\simeq \frac{\Phi _{0}/c}{{%
2.3nkT}_{e}}\approx U_{0}\frac{\left\langle {\varepsilon }_{{\rm {%
\scriptscriptstyle EUV}} }\right\rangle }{2.3kT_{e}}  \label{eqnXi}
\end{equation}
where $k$ is Boltzmann's constant, $c$ the speed of light, $n$ and $T_{e}$
are the hydrogen density and the electron temperature at the irradiated
cloud surface (indicated by a subscript zero elsewhere). We define $S_{*}={%
\Phi }_{0}{/}\left\langle {\varepsilon }_{{\rm {\scriptscriptstyle EUV}}
}\right\rangle =\int \frac{{\varphi _{0}(\nu )}}{h\nu }{d\nu }$ to be the
ionizing photon flux (cm$^{-2}$s$^{-1}$) impinging at the surface such that $%
U_{0}=S_{*}/n c$ is the canonical ionization parameter. $\left\langle {%
\varepsilon }_{{\rm {\scriptscriptstyle EUV}}}\right\rangle $ ($=\langle
h\nu_{{\rm {\scriptscriptstyle EUV}}}\rangle$) is the local average EUV
photon energy.

In hydrostatic equilibrium, the gas pressure gradient must match the local
radiative volume force exerted by photon absorption, and this force can be
separated into its two components, the force acting on the gas through
photoionization, and the component due to dust absorption;
\begin{equation}
\begin{align}
F_{{\rm rad}}(x)=&\sum_{m=1}{\sum_{i=0}{n_{i}(X_{m}^{+i})\int_{\nu
_{m}^{i}}^{\infty }{\frac{{\varphi (\nu )}}{{c}}}a_{\nu }(X_{m}^{+i})~d\nu }}
\nonumber \\
&+n_{H}\int_{\nu _{m}^{k}}^{\infty } \left[\kappa_{{\rm abs}}\left( \nu
\right)+\kappa_{{\rm sca}}\left ( \nu \right)
\left(1-g\left(\nu\right)\right)\right]{\frac{{\varphi (\nu )}}{{c}}}d\nu
\label{eqnforce1}
\end{align}
\end{equation}
where $\varphi (\nu )$ is the local radiation flux at frequency $\nu $, $%
n_{i}(X_{m}^{+i})$ the number density of ion $+i$ of atomic species $m$, $%
a_{\nu }(X_{m}^{+i})$ the corresponding photoionization cross section with
threshold $\nu _{m}^{i}$. In the dust absorption term, $\kappa_{{\rm abs}%
}\left( \nu \right) $ and $\kappa_{{\rm sca}}\left( \nu \right) $ is the
dust absorption cross section and scattering cross section respectively,
normalized to the hydrogen density and $g\left(\nu\right) = \left\langle
\cos \theta _{\nu}\right\rangle $ is the mean scattering angle of the dust
grains at this frequency. The boundary condition is given by $P_{0}$, the
pressure at the irradiated face.

We can readily estimate a critical value of $U_{0}$ above which dust
absorption dominates over the photoelectric absorption. For an equilibrium
photoionized plasma, the absorption of ionizing photons in a slab is simply
equal to the local recombination rate in the plasma:
\[
dS_{*}/dx=-\alpha _{B}n^{2}
\]
where $\alpha _{B}$ is the recombination rate to excited states of hydrogen.
The absorption of ionizing photons by dust is given by:
\[
dS_{*}/dx=-\kappa nS_{*}
\]
where $\kappa $ is the effective dust opacity of order $\sim 10^{-21}\,{\rm %
cm}^{2}$ per H atom. Dust absorption becomes relatively more important as
the strength of the ionizing field increases, and dust becomes the dominant
opacity in the plasma when:
\begin{equation}
U_{0}>\frac{\alpha _{B}}{c\kappa }  \label{Udom1}
\end{equation}
For a typical value of $\alpha _{B}=2\times 10^{-13}\,{\rm cm}^{3}{\rm s}%
^{-1}$, the critical ionization parameter is $\sim 0.007$.

This value is closely related to that at which radiation pressure starts to
dominate either the gas pressure or the dynamical acceleration of the
plasma. If $\left\langle \varepsilon _{{\rm {\scriptscriptstyle EUV}}%
}\right\rangle $ is the mean energy of the ionizing photons absorbed by gas,
and $\left\langle \varepsilon _{{\rm {\scriptscriptstyle FUV}}}\right\rangle
$ is the mean energy of the ionizing absorbed by dust, equation (\ref
{eqnforce1}) can be approximated by:
\begin{eqnarray}
F_{{\rm rad}}(x) &=&\frac{1}{c}\alpha _{B}n^{2}\left\langle \varepsilon _{%
{\rm {\scriptscriptstyle EUV}}}\right\rangle +\frac{1}{c}S_{*}\left\langle
\varepsilon _{{\rm {\scriptscriptstyle FUV}}}\right\rangle \kappa n
\nonumber \\
&=&\frac{n^{2}\left\langle \varepsilon _{{\rm {\scriptscriptstyle EUV}}%
}\right\rangle }{c}\left[ \alpha _{B}+\frac{\xi }{\psi }U_{0}\kappa c\right]
\label{eqnFaprx}
\end{eqnarray}
where $\left\langle \varepsilon _{{\rm {\scriptscriptstyle EUV}}%
}\right\rangle /\left\langle \varepsilon _{{\rm {\scriptscriptstyle FUV}}%
}\right\rangle =\psi $ , and the effective ionization parameter for the
dust--absorbed photons $U_{{\rm {\scriptscriptstyle FUV}}}=\xi U$. Generally
speaking, both $\xi $ and $\psi $ will be greater than unity. Clearly, dust
becomes the dominant term in this equation when:
\begin{equation}
U_{0}>\frac{\psi \alpha _{B}}{\xi c\kappa }  \label{Udom2}
\end{equation}
this expression is the same as equation (\ref{Udom1}) except for the ratio $%
\psi /\xi $ , which is of order unity. In the typical ionization field
expected around an AGN, ${\varphi _{0}(\nu )\propto \nu }^{-1.4}$and with a
`standard' dust model with solar abundances, a \citet{MRN77} (MRN) size
distribution $n(a)\propto a^{-3.5}$ and a lower size cutoff of 10 nm , we
estimate that this condition gives a critical ionization parameter $%
U_{0}\sim 0.01.$

If the ionization parameter into a static plane-parallel slab of ionized gas
is increased above this critical value, then the radiation pressure gradient
imposed as the EUV radiation field is absorbed through the ionized slab will
induce a corresponding local gas pressure gradient. Thus, if the initial
(outer layer) pressure of the ionized gas is $P_{0}$ and the radiation
pressure in the EUV field is $P_{rad}$ then the gas pressure close to the
ionization front is $P_{0}+P_{rad}.$ Thus, at high ionization parameter, the
pressure in the ionized gas close to the ionization front is {\it determined
}by the externally imposed ionization parameter, $U_{0},$ and the {\it local
}ionization parameter becomes independent of the initial ionization
parameter. In this way, the emission line spectrum in the low- and
intermediate- ionization species will become effectively independent of the
external ionization parameter. This is the core of our explanation of why
the emission line spectra of NLR are so similar one to another.

While $U_{0}$ defines the ionization parameter at the face of the cloud,
one can define a {\it mean} ionization parameter $\bar{U}=S_{*}/\bar{n}%
c=U_{0}n_{0}/\bar{n}$ where $\bar{n}$ is simply the mean density, that is an
emissivity averaged density using a weight $\propto n_{e}^{2}$. %
\citet{bwrs97} showed that $\bar{U}$ tends asymptotically towards a constant
value in the limit of high $U_{0}$.

\section{Results of Models}

We have constructed static, plane-parallel, radiation-pressure dominated,
dusty photoionization models using the code MAPPINGS\ III, which includes
the implementation of eqn. (\ref{eqnforce1}) to evaluate the local radiation
pressure gradient and hence the local density gradient. MAPPINGS\ III
~treats several of the properties of dust such as photoelectric heating, as
described previously by \citet{do:su00}. The dust model includes both
silicaceous and carbonaceous grains with a MRN grain size distribution and
uses the absorption and scattering data of \citet{LaorDr93} (using the more
recent smoothed UV data for silicates)\footnote{%
Available at http://www.astro.princeton.edu/$\sim$draine/dust/dust.diel.html}.
Here we have chosen the simplest ``standard'' values for the input
parameters. For example, the models all used the \citet{anders89} solar
abundance set, with the depletion factors for each element given in Table 1.
The input ionizing spectrum for the following models was a standard powerlaw
continuum with ${\varphi _{0}(\nu )\propto \nu }^{-1.4}$.The effect of
varying these and other input parameters will be investigated in a following
paper (Groves et al.~(2001) in preparation).

To facilitate comparison, and to separate the effects of internal density
structure, radiation pressure and photoelectric heating by the grains, we
have run four families of models with increasing ionization parameter $U_{0}$
(defined at the outer boundary of the ionized slab). In order of increasing
physical sophistication (and reality) these are:

1. Isochoric (constant density) dust-free models with heavy element
abundances in the gas phase which match those of the models including dust.

2. Isobaric dust-free models. Again these have gas-phase heavy element
abundances matching the models with dust. These models are isobaric in the
sense that the product $nT$ is kept constant, but the radiation pressure is
not explicitly included.

3. Isochoric dusty models.

4. Isobaric dusty models without radiation pressure. These provide a
dusty comparison to the complete model which shows the difference that
radiation pressure makes. 

5. Complete isobaric dusty models, which include the effect of radiation
pressure on the density structure of the ionized gas.

To avoid differential effects due to different electron density in the
low-excitation cases, in all cases the initial density was adjusted so that the density of the region
emitting in the [S\thinspace {\sc ii}] lines was maintained constant at
about 550 cm$^{-3}$ to within 10\%. This density was chosen to be typical of the electron
densities inferred from the [S\thinspace {\sc ii}] $\lambda \lambda 6717/6731
$ ratio in the NLRs of Seyfert Galaxies. The models were terminated at a
point where the gas temperature dropped rapidly to below 5000K and the
emission in the optical forbidden lines is quenched.

\begin{figure}
\scalebox{0.8}[0.8]{\plotone{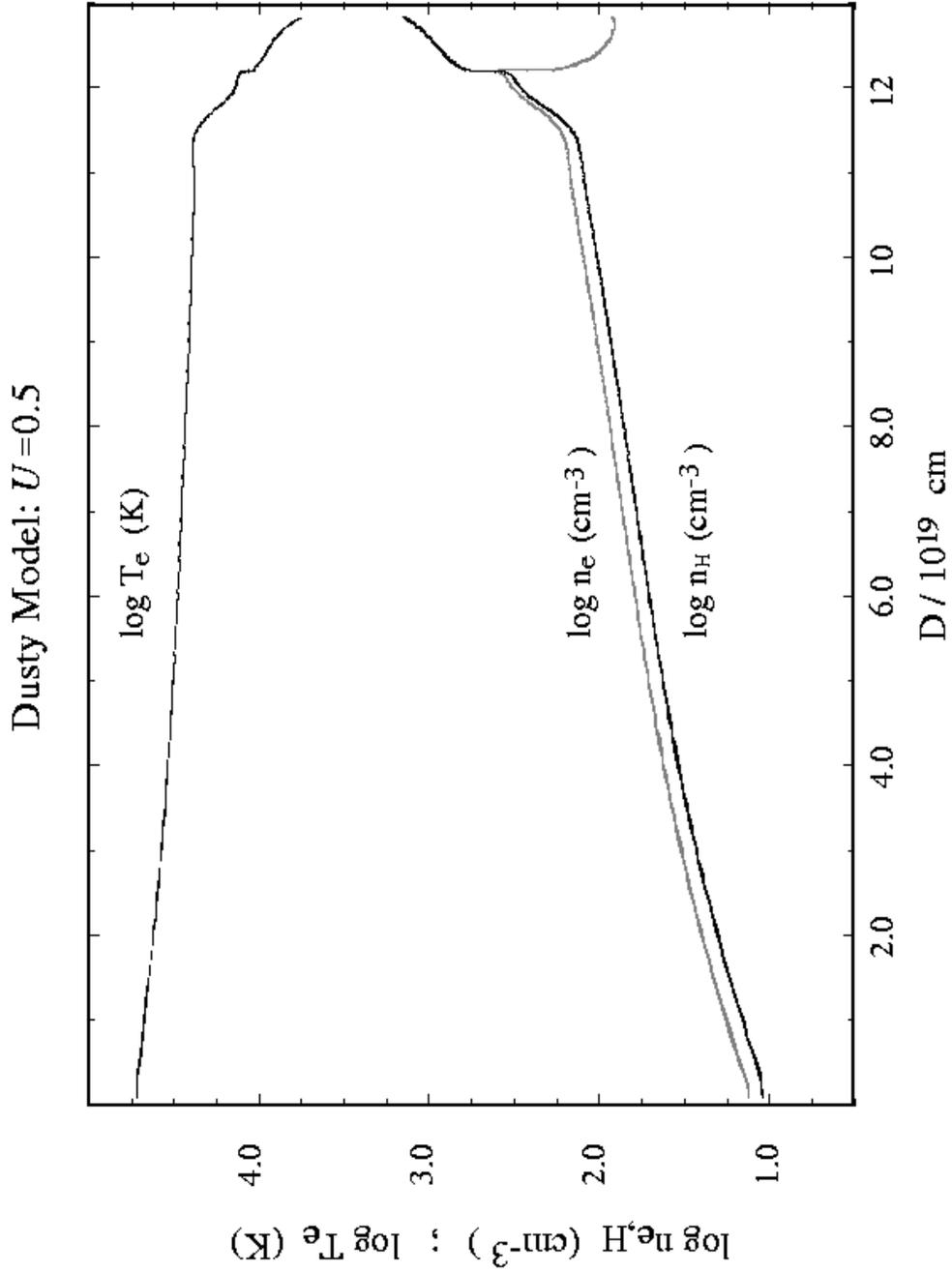}}
\caption{\label{structure}Temperature and density structure in a
radiation-pressure dominated, dusty model with $U_0=0.5$}
\end{figure}

Generally in the literature we encounter only the first or the
third of these sets of models. The profound difference in the internal
structure of the isobaric dusty models including radiation pressure compared
with a ``standard'' isochoric model is emphasized in Fig.~\ref{structure}
which shows the internal density and temperature structure produced in the
dusty, radiation pressure dominated slab model with ionization parameter of 
$U_{0}=0.5.$ In this model, the radiation pressure in the EUV radiation field is
roughly 30 times the initial gas pressure. At the ``front'' surface of the
cloud the opacity of the cloud is entirely dominated by dust, and the
photoelectric heating by dust is very large. As a consequence, the electron
temperature in this zone is very high, of order 25000K. The electron
temperature decreases strongly throughout the model towards the ionization
front, and this steepens the density gradient in the model.

This strong density gradient which is set up between the high
ionization and low ionization regions can provide a physical explanation of
the pressure difference between these regions which was required in the $%
A/M_{{\rm I}}$ models of \citet{bws96}. As expected, the pressure in the
low-ionization region scales as the radiation pressure at high $U_{0}.$
Thus, the local ionization parameter characterizing the low and intermediate
ionization region of the model becomes independent of $U_{0}.$

The emission line ratios produced by the various classes of model can be
quite different one from another. We have chosen to display here only those
which provide a good separation between the different classes of model, and
so have a diagnostic utility. Any of the of line emission ratios used by %
\citet{VO87} will give such a separation.

\begin{figure}
\plotone{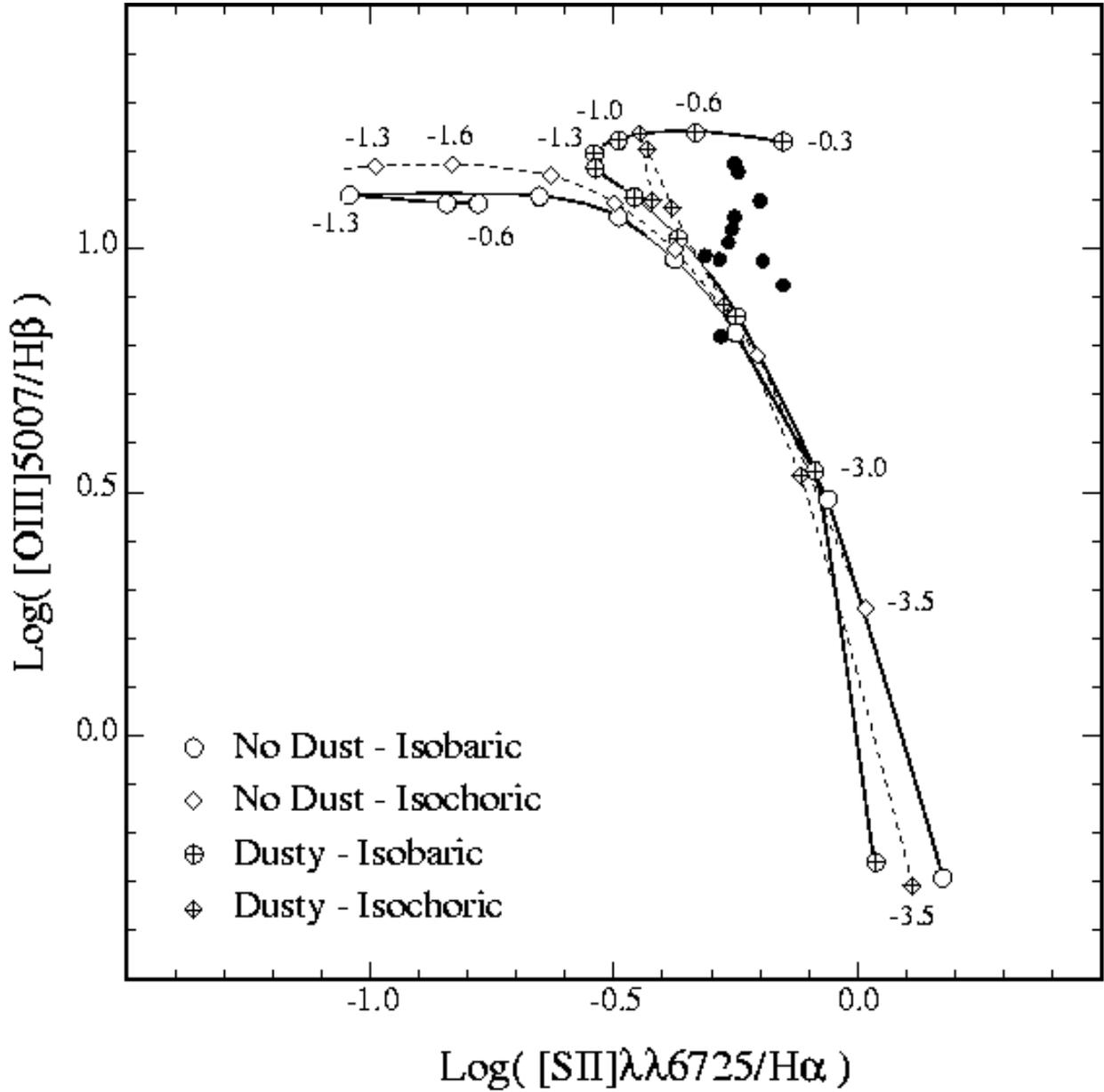}
\caption{\label{SII-OIII} Line diagnostic diagram of [\oiii]$\lambda
5007/$H$\beta$ vs. [\sii]$\lambda\lambda 6717,31/$H$\alpha$ showing the output
from four families of models. These are split into  two dust free models
(empty markers) and
two dusty models (crossed markers), with an isochoric model (dashed lines) and an
isobaric model (solid lines) in each case. The isobaric, dusty model also
includes the effect of radiation pressure. Along each model line marks
indicating log$U_{0}$ are shown. The solid circles are observations from
\citet{Allen99}. Note that the dust free cases do not match the observations
whereas both dusty models provide reasonable solutions.}
\end{figure}

In figure \ref{SII-OIII} we plot one of these diagnostic diagrams, the
[S\thinspace {\sc ii}] $\lambda \lambda 6717+6731$\AA /H$\alpha $ ratio $vs.$
the [O\thinspace {\sc i\thinspace ii}] $\lambda 5007$\AA /H$\beta $ ratio.
The observational points are from \citet{Allen99}. For the dust-free case,
both the isochoric nor the isobaric sequence can reproduce the observed data,
though the fit would be improved with a flatter spectral
index.
The closest fit is found in the region $\log U_{0}\sim -2.8$. As has been shown by
many other authors, only a very restricted range of ionization parameter
provides a reasonable fit.

For the dusty models, both the isochoric and isobaric sequences approach the
region of the observed NLR at high ionization parameter. The
dusty, radiation pressure affected isobaric model shows very slow variation in
the line ratios with $U$ at high ionization parameter and these line ratios are
in the region of
interest. The dusty, isochoric models actually begin to head away from the
observations as we increase in ionization parameter.

\begin{figure}
\plotone{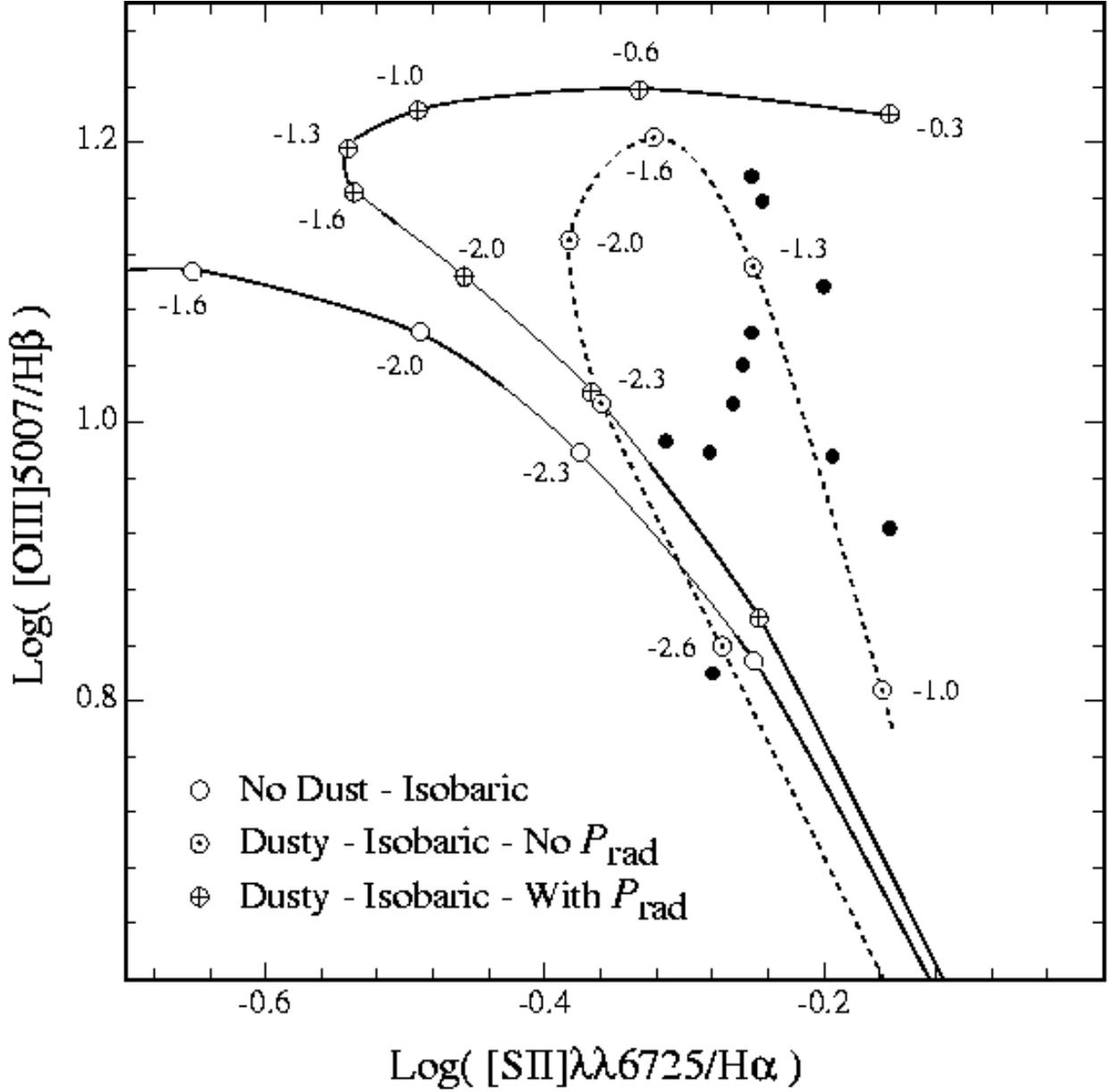}
\caption{\label{IsoBmodels} A close-up of the region of interest in figure
\ref{SII-OIII} showing only the isobaric models. As well as the dust free
isobaric model (empty circles) and the dusty, radiation pressure affected
model (crossed circles) from the previous figure, a dusty model without the
effects of radiation pressure is included (dashed line with dotted circles).
Note the model without $P_{rad}$ does match the observations well, but the
line ratios do not stagnate with changes in $U$ like the complete dusty model.}
\end{figure}

Figure~\ref{IsoBmodels} is a close up of the region of interest, showing
only isobaric models. In addition to the two isobaric models described
above, a dusty, isobaric model without radiation pressure is included for comparison.
This figure illustrates the self-regulating effect of the
radiation pressure in producing nearly constant line ratios for 
$U_{0}>10^{-2}$. While the radiation pressure free case does reproduce the
observed data quite well, it does so only for a small range of ionization
parameter, $-2.3<\log U_{0}<-1.3$. By $\log U_{0} \sim -1.0$ the model
has left the region in which the observations lie. The complete dusty models, which
include the effects of radiation pressure, not only lie in the region of the observed
data but also display line ratios which
stagnate with respect to changes in $U$, over a broad range of values. 
The implication of this is that, provided $U_{0}\gg 10^{-2}$
, a cloud of given density could be moved radially by a large amount without
producing any appreciable change in its excitation state. Such a behavior
is frequently seen, and is interpreted in the ``classical'' photoionization
models as indicating a constant ionization parameter.

\begin{figure}
\plotone{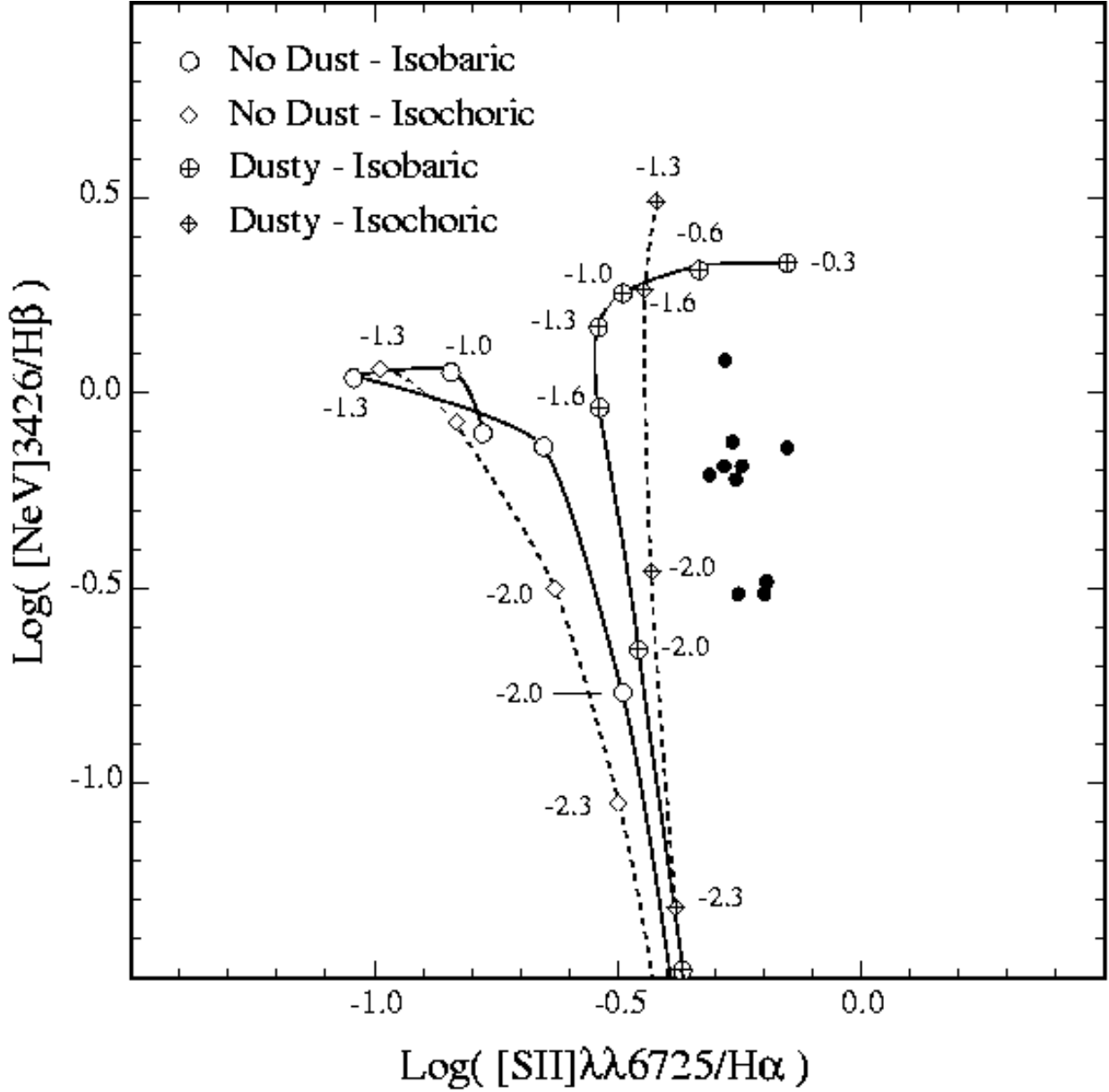}
\caption{\label{SII-NeV} Line diagnostic diagram of [\nev]$\lambda
3426/$H$\beta$ vs. [\sii]$\lambda\lambda 6717,31/$H$\alpha$ with the same
symbols as figure \ref{SII-OIII}. Note that in consideration with
\ref{SII-OIII} that only the dusty models provide reasonable solutions the
isochoric case with $-2.0<\log U_{0}<-1.2$ and the isobaric case with
$-1.6<\log U_{0}<0$.}
\end{figure}

The conclusions derived from figure \ref{SII-OIII} and \ref{IsoBmodels} are reinforced when we
replace the [O\thinspace {\sc i\thinspace ii}] line by the higher-excitation
[Ne\thinspace {\sc v}] line. This is shown in figure \ref{SII-NeV}. Now the
range of acceptable models is appreciably reduced. From 
inspection of both Figures \ref{SII-OIII} and \ref{SII-NeV} we see that the
dusty 
models that include radiation pressure provide acceptable solutions in the
range $-1.6<\log U_{0}<0.$ The acceptable range of solutions for the dusty
isochoric models is much more restricted; $-2.0<\log U_{0}<-1.2,$ and there
are no really acceptable solutions for {\em any} dust-free model. This
illustrates a major advantage of the radiation pressure dominated model in
that it encompasses a wider range of ionization species than any simple
isochoric or isobaric model. The reason for this is that as density increases
inward, one is effectively spanning a very wide range in the local
ionization parameter. Furthermore, the extra heating provided by dust
photoheating scales with $U_{0}$ and the higher electron temperatures
enhance the intensities of the high ionization species in a way that the dust 
free models cannot account for. This effect is further enhanced by
dust absorption which strongly reduces the luminosity of the hydrogen
recombination lines in the high ionization zone. At the same time, it
enhances the He\thinspace {\sc ii}~$\lambda 4686$\AA /H$\beta $ ratio to
values as high as 0.5-0.6 which is in better agreement with observation, and
similar to the value obtained in the matter-bounded models of \citet{bws96}.

\begin{figure}
\plotone{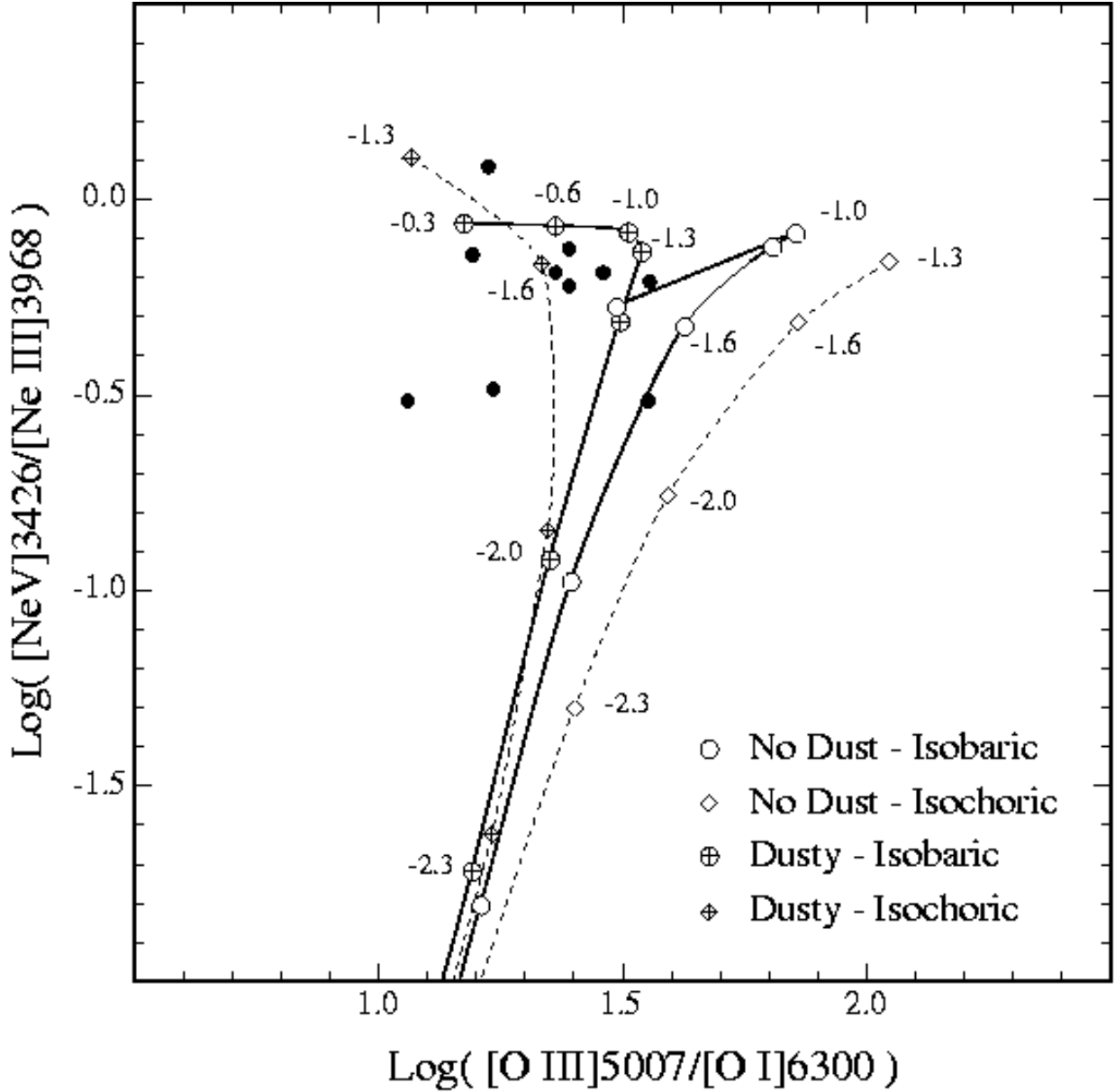}
\caption{\label{excitation} Line diagnostic diagram of [\nev]$\lambda
3426/ $[\neiii]$\lambda 3968$ vs. [\oiii]$\lambda 5007/$[\oi]$\lambda 6300$,
with the same symbols as figure \ref{SII-OIII}. Note that these line ratios are
sensitive to excitation, and with inspection of this and the previous two
figure it can be seen that again the isobaric dusty model provides a
reasonable solution between $-1.6<\log U_{0}<0$.}
\end{figure}

If we look only at line ratios which are sensitive to the excitation, then
dust-free models might seem to provide an acceptable fit. An example is
shown in figure \ref{excitation} which plots the [O\thinspace {\sc %
i\thinspace ii}]/[O\thinspace {\sc i}] line ratio against the [Ne\thinspace
{\sc v}]/[Ne\thinspace {\sc iii}] ratio. Here similar values of the
ionization parameter produce very similar values of the [Ne\thinspace {\sc v}%
]/[Ne\thinspace {\sc iii}] ratio, while the [O\thinspace {\sc i\thinspace ii}%
]/[O\thinspace {\sc i}] line ratio is more strongly affected by the
temperature in the [O\thinspace {\sc i}] - emitting zone. Note that, again
the dusty models including radiation pressure provide acceptable solutions
in the range $-1.6<\log U_{0}<0.$

\begin{figure}
\plotone{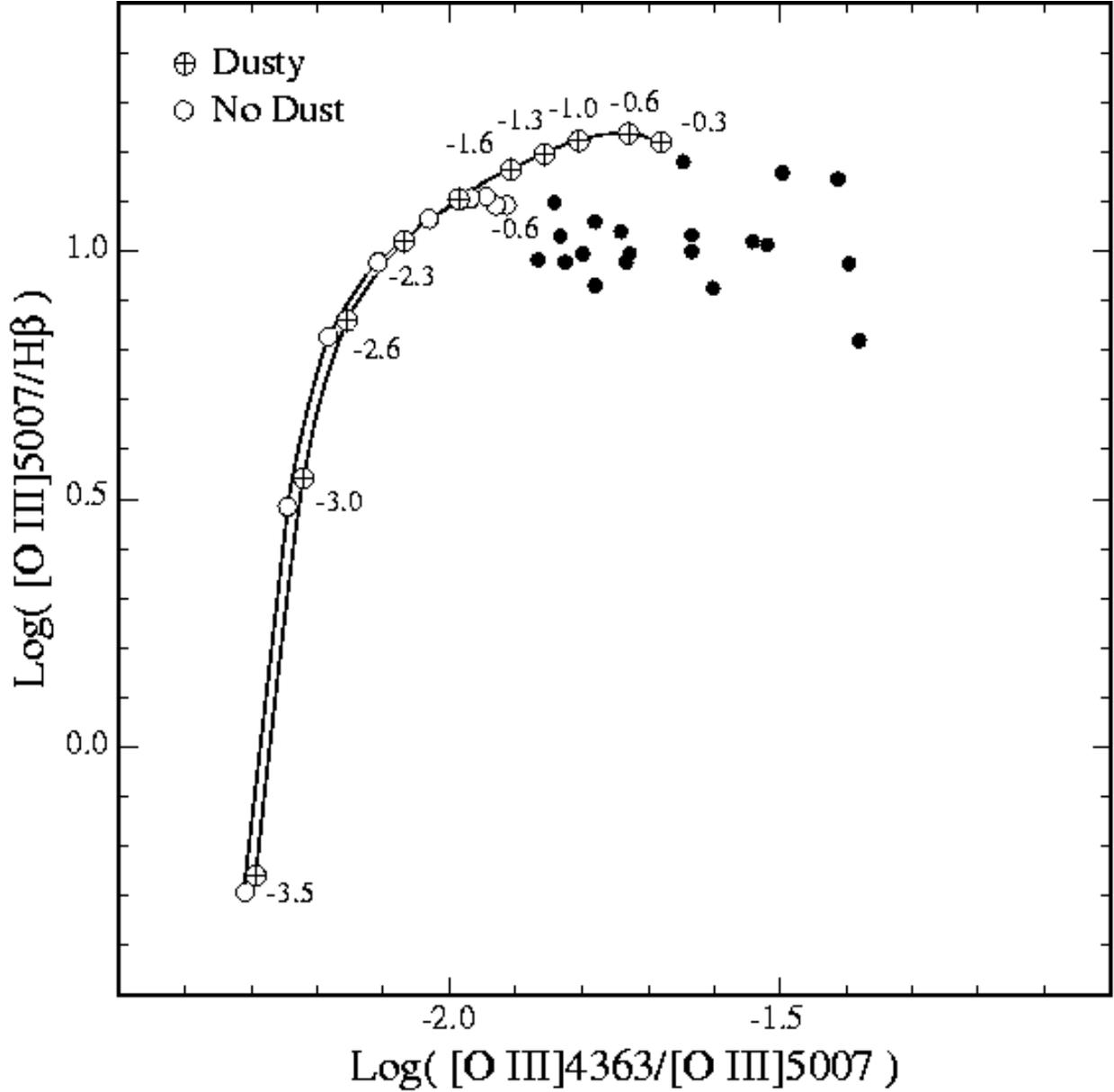}
\caption{\label{temperature} Line diagnostic diagram of [\oiii]$\lambda
5007/$H$\beta$ vs. [\oiii]$\lambda 4363/\lambda 5007$ with two cases: a dust
free isobaric model (empty circles) and a dusty, isobaric with radiation pressure
model (crossed circles). The marks on the model lines indicate $\log U_{0}$. The
solid circles are observations drawn from both \citet{Allen99} and
\citet{Tad89}. Note that with the consideration of figures
\ref{SII-OIII}--\ref{temperature} that the isobaric, dusty, radiation pressure
affected models provide the most acceptable solutions with $-1.6<\log U_{0}<0$.}
\end{figure}

The effect of photoelectric heating is best investigated in terms of the
electron temperature in the [O\thinspace {\sc i\thinspace ii}] - emitting
region. This is shown in figure \ref{temperature} which plots the
temperature sensitive [O\thinspace {\sc i\thinspace ii}] $\lambda 4363$\AA
/5007\AA  ratio against [O\thinspace {\sc i\thinspace ii}] $\lambda 5007$\AA
/H$\beta .$ For clarity we have omitted the isochoric models, since these
largely overlap with the isobaric cases. The observed points are drawn from
both \citet{Allen99} and \citet{Tad89}. The dust-free models are clearly
much cooler than the dusty models and fail to reproduce the observed
[O\thinspace {\sc i\thinspace ii}] temperatures for any ionization
parameter. This is the famous ``temperature problem'' of NLR, and which was
discussed in detail by \citet{Tad89}. Photoelectric heating by dust,
proposed as the solution to a similar problem occurring in planetary nebulae
by \citet{do:su00} seems to provide enough additional heating in NLR to
bring the predicted electron temperatures into the observed range. Once
again, for dusty isobaric radiation-pressure dominated models, acceptable
models for the [O\thinspace {\sc i\thinspace ii}] $\lambda 4363$\AA /5007\AA 
~ratio lie in the range $-1.6<\log U_{0}<0.$

\begin{figure}
\plotone{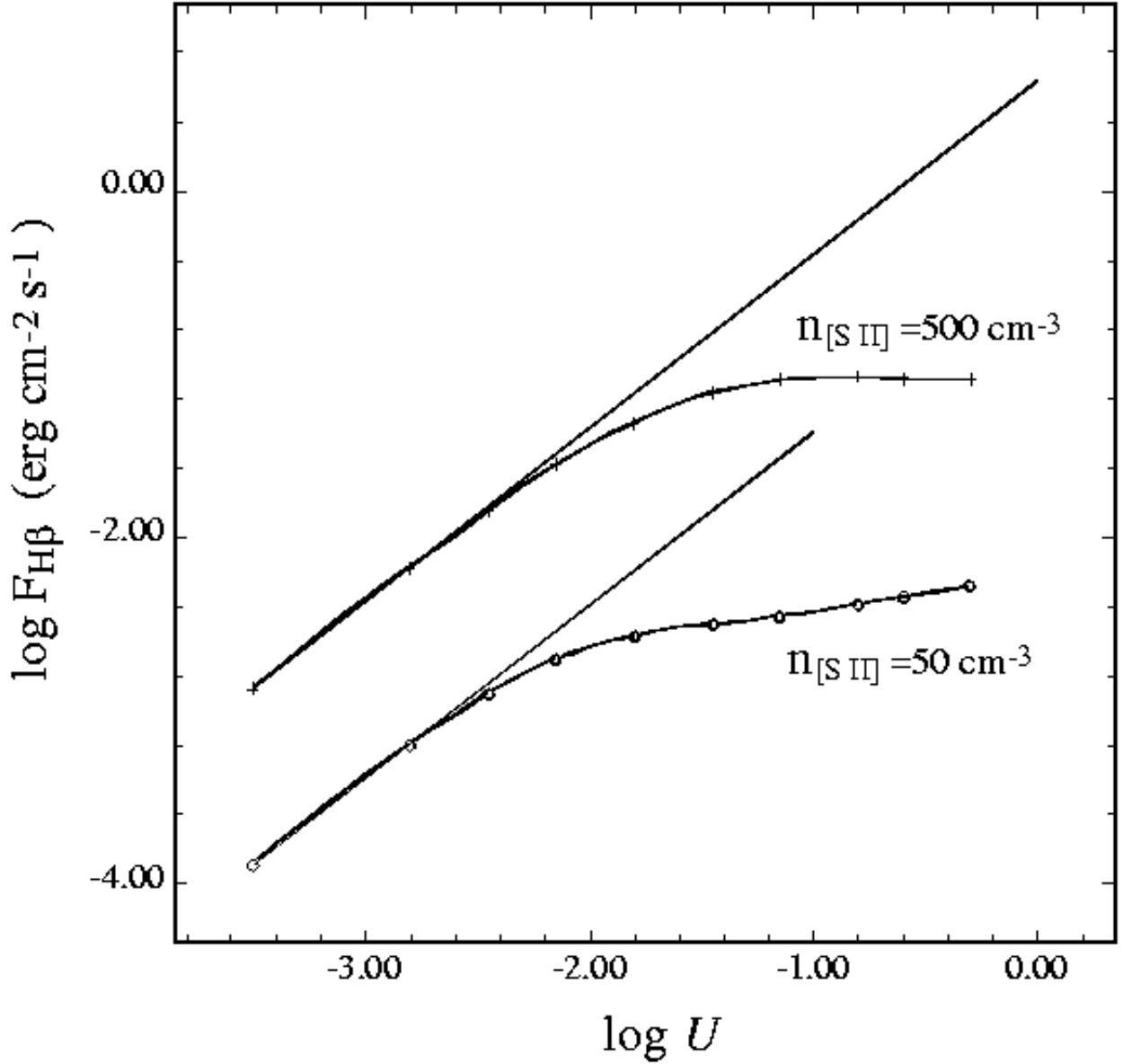}
\caption{\label{Hbflux} H$\beta$ flux vs. ionization parameter, showing two 
groups of models. The two groups shown differ in density, with
the crosses indicating a density at the back of the cloud of $\sim 550$
cm$^{-3}$ and the circles, $\sim 55$ cm$^{-3}$ (as labelled).
Within each group the
thin lines show a 'standard' (isochoric) model without dust, and the thick lines 
the result from our radiation-pressure dominated, dusty model, showing the
limiting of the H$\beta$ flux.  }
\end{figure}

A final major effect of the addition of dust in these models is to limit the
H$\beta $ surface 
brightness as a consequence of the absorption of ionizing photons by
dust. Since dust absorption 
strongly dominates over the gas absorption in the regime of high ionization
parameter, this limits the ionized column (Str$\ddot{\textrm{o}}$mgren column) to a
maximum value. For a fixed density or pressure
in the region emitting the [S II] lines, the surface brightness of the
ionized zone in H$\beta $ saturates at a maximal value at high $U_{0}$. By
contrast, at low $U_{0},$ dust absorption is negligible compared with gas
absorption, and the H$\beta $ flux scales with the ionization parameter.
This behavior is shown in figure \ref{Hbflux}.

If the [S II] density is lowered, then, at at a given $U_{0}$ the H$\beta $
flux simply scales with the density in the low-$U_{0}$ regime. However,
since the dust and the gas absorption terms have a different density
dependence, at a given $U_{0}$ in the high-$U_{0}$ regime, a lowering of the
density in the [S II] zone leads to a greater fraction of the incident
radiation field being absorbed by dust. The result of this is that the
saturation value of the H$\beta $ surface brightness is reduced by a greater
factor than simply the ratio of the densities. The roll-off between the low-$%
U_{0}$ and high-$U_{0}$, regimes must therefore occur at a slightly lower
value of $U_{0}$ than in the high density case.

The plane-parallel models presented here strictly apply only along the
(stagnation) line which lies directly towards the ionizing source on the front
face of a photoevaporating cloud. Towards the sides of the cloud, lateral
motions will be developed in the ionized plasma by the transverse gas pressure
gradient. Around the cloud edges, the radiation pressure force acts radially
outwards from the central engine. According to equation (\ref {eqnforce1}), this
drives an accelerating flow radially away from the central AGN. Such a
radiatively-driven outflow was considered by \citet{Bin98} in the context of
the coronal gas. In the case of NGC\ 1068, new HST STIS data presented by
\citet{Cecil01} shows [O\thinspace {\sc iii}] tails of clouds about 100 pc out
from the nucleus being accelerated up to 2000 km s$^{-1}$ or greater, over a
distance of about 0.15 arc sec (10 pc). This corresponds to an acceleration of
order $6\times 10^{-4}$cm~s$^{-2}$. Let us now examine whether this 
acceleration
could be due to radiation pressure.

According to the observations of \citet{KC2000}, the density given by the
[S\thinspace {\sc ii}] lines in NGC 1068 is about 2500 cm$^{-3}.$ In the high ionization
parameter models, this density is determined entirely by the radiative 
pressure, and such
a pressure is obtained from a central source giving a surface luminosity of
$\sim 1500$ erg cm$^{2}$ s$^{-1}$ at the outer boundary of the photoablating cloud.
This figure is in excellent agreement with that inferred from the bolometric
FIR luminosity 
(\citet{Do98}) $\sim 1.5\times 10^{11}$ $L_{\odot }$. This gives a flux at a
distance of 70 pc from the nucleus (typical for these clouds) of about
1700 erg cm$^{2}$ s$^{-1}.$ 

If these radiative forces go into
accelerating the gas in the ionized tails of the NGC1068 clouds, then from
equation (\ref{eqnFaprx}) we would require a flux of order $1.5\times 10^{4}$
erg cm$^{-2}$ s$^{-1}$ to provide the observed acceleration. The inferred flux
is a factor ten lower than this, and this would produce a final outflow 
velocity of
order 700km s$^{-1}$. The only way around this problem would be to increase the
force per unit mass of gas by increasing opacity of the dust. This 
could be done
through grain shattering, which according to \citet{Jones96} will 
occur whenever
grain-grain collision velocities are greater than the sound speed in the grain
material, about 2 km s$^{-1}$. This could be achieved with either subsonic
turbulence in the gas, or else subsonic radiatively-driven grain drift
velocities.

A potential mechanism for grain shattering is as follows. Close to the
ionization front, and within it, the net grain charge is negative, being
dominated by collisions. When the grains are charged their motion is strongly
coupled to the gas via Coulomb interactions. As the grains are advected with
the gas flow from the ionization front, the local radiation field becomes
stronger and the charge on the grains becomes smaller as photoelectric charging
becomes more important. Eventually, the net charge on an individual grain
becomes zero as the collisional and photoelectric charging exactly balance. At
this point the grain can decouple from the gas flow and radiative pressure
forces push it back towards the ionization front. The terminal drift velocity
that the grain can reach is determined by the gas density, $\rho$, and the
radiation field intensity, $F$; $\textrm{v}_{\textrm{drift}} = (F/c\rho)^{1/2}$. For
the parameters computed for NGC 1068 this could  amount to $\sim 200$
\kms. However, long before this velocity is reached, the grain will become
charged once more, and the grain motion is again locked to the gas
motion. Thus, when their charge becomes zero, the grains will tend to
oscillate about the position where this happens until another (charged) grain
strikes and shatters them. The shattered fragments have their equilibrium
point further out in the ionized flow, and of course they are much more
vulnerable to the other grain destruction mechanisms we have mentioned. In our
models for NGC 1068 grains are completely shattered in the region which
produces [\nev] and [\fevii].

The idea that the ionized tails of the cloudlets in NGC~1068 are driven by
radiation pressure acting on the shattered dust seems a reasonable one, but it is clear
that the full solution of this problem will require a hydrodynamic model which
explicitly includes the grain physics and destruction processes. We intend to
present such a model in a future paper.

If dust is indeed destroyed at some point in the outflow, the gas will continue to flow
outward from the cloud, but radiation pressure effects become relatively
ineffective in determining the dynamics. In this case the cloud will become
surrounded by a highly ionized coronal region with an equilibrium electron
temperature of order $8\times 10^{4}$K. On the basis of continuity of pressure,
we estimate that in the case of NGC1068, the density of this region is $10-20$
cm$^{-3}$ and therefore that the effective ionization parameter is of order
unity, or somewhat greater. This would be sufficient to explain the extreme
degree of ionization observed in this coronal region (\citet{KC2000}).

Finally, let us examine if the hypothesis that dust can be accelerated in the
cometary tail of the cloud is also consistent with the hypothesis that dust can also
be destroyed in entering the coronal region. From the acceleration in the tail
($6\times 10^{-4}$cm~s$^{-2}$) and the length of the tail (10pc) we 
can estimate
the transit time of material through the tail; $10^{4}$yr. Along the stagnation
line in the direction of the source, material flows off the cloud at about
the sound speed, 20 km s$^{-1}$. However, the photoionization model for NGC 1068
shows that the thickness of the ionized zone is about $10^{18}$cm. Therefore,
the dusty plasma flows through this region in about $1.6\times 10^{4}$yr. These
timescale are quite comparable, which strongly suggests that the length of the
[O\thinspace {\sc iii}] tails and the production of the coronal region in NGC
1068 are both moderated by the dust destruction, occurring on a timescale of
order $10^{4}$yr.

\section{Conclusions}

In this paper we propose that the NLR and ENLR of Seyfert galaxies can be
self-consistently modelled by radiation-pressure dominated dusty
photoionized regions surrounding photoablating dense clouds. By using the
code MAPPINGS\ III, we have shown that these display a strong density
gradient increasing towards the ionization front, are characterized by
strong photoelectric heating by the dust and in which the dominant
absorption of the ionizing continuum is due to the dust. In these models,
the gas pressure close to the ionization front is determined by the
radiation field. The models provide:

\begin{itemize}
\item  A line spectrum consistent with the observations, and which remains
approximately invariant against changes in $U_{0}$\ that span a range larger
than 2 orders of magnitude.

\item  A higher electron temperature than can be achieved in standard
photoionization models, and which could help explain the long standing
``temperature problem'' of NLR and ENLR.

\item  An H$\beta $ surface brightness which reaches a maximum value at high
values of the initial ionization parameter $U_{0}$.
\end{itemize}

In addition, they may also offer an explanation for:

\begin{itemize}
\item  The co-existence of the low-ionization zones with a extremely high
ionization parameter coronal region.

\item  Strong radial radiative acceleration of dusty gas which may explain
the broad and blue-shifted [O\thinspace {\sc iii}] profiles seen in many
Seyfert galaxies.
\end{itemize}

We will systematically explore the properties of such models in future
papers.

\begin{acknowledgements}
The work of LB was supported by the CONACyT grant 32139-E and was
carried out while a Visitor at the RSAA supported by a RSAA visitors grant. 
M. Dopita acknowledges the support of the Australian National
University and the Australian Research Council (ARC) through his
ARC Australian Federation Fellowship, and also under ARC
Discovery project DP0208445. The authors
would like to thank Ernesto Oliva for his many helpful comments and for pointing
out an error that the authors missed. 
\end{acknowledgements}

\newpage

\begin{deluxetable}{lrr}
\tabletypesize{\small}
\tablecaption{Solar metallicity ($Z_{\odot}$) and depletion factors (D) adopted
for each element.\label{Z_table}}
\tablehead{
\colhead{Element}
& \colhead{$\log({\rm Z_{\odot}})$}
& \colhead{$\log({\rm D})$}\\
}
\startdata
H & 0.00 & 0.00 \\
He & -1.01 & 0.00 \\
C & -3.44 & -0.30 \\
N & -3.95 & -0.22 \\
O & -3.07 & -0.22 \\
Ne & -3.91 & 0.00 \\
Mg & -4.42 & -0.70 \\
Si & -4.45 & -1.00 \\
S & -4.79 & 0.00 \\
Ar & -5.44 & 0.00 \\
Ca & -5.64 & -2.52 \\
Fe & -4.33 & -2.00 \\
\enddata
\end{deluxetable}

\clearpage
\onecolumn

\begin{thebibliography}
\singlespace
\bibitem[Allen et al.(1999)]{Allen99}  Allen, M. G., Dopita, M. A.,
Tsvetanov, Z. I., \& Sutherland, R. S. 1999, \apj, 511, 686

\bibitem[Anders \& Grevasse(1989)]{anders89}  Anders, E. \& Grevesse, N., %
\gca, 1989, 53, 197

\bibitem[Baldwin et al.(1991)]{Baldwin91}  Baldwin, J.~A., Ferland, G.~J.,
Martin, P.~G., Corbin, M.~R., Cota, S.~A., Peterson, B.~M., \& Slettebak,
A.\ 1991, \apj, 374, 580

\bibitem[Binette(1998)]{Bin98}  Binette, L., 1998, \mnras, 294, L47

\bibitem[Binette et~al.(1997)]{bwrs97}  Binette, L., Wilson, A. S., Raga, A.
\& Storchi-Bergmann, T., 1997, \aap, 327, 909

\bibitem[Binette et~al.(1996)]{bws96}  Binette, L., Wilson, A. S.,
Storchi-Bergmann, T., 1996, \aap, 312, 365

\bibitem[Bottorff et al.(1998)]{Bott98}  Bottorff, M., Lamothe, J., Momjian,
E., Verner, E., Vinkovi\'{c} , D., \& Ferland, G.\ 1998, \pasp, 110,
1040

\bibitem[Cecil et al.(2001)]{Cecil01}  Cecil, G., Dopita, M. A., Groves, B.,
Ferruit, P., P\'{e}contal, M., \& Binette, L., 2001, \apj, submitted

\bibitem[de Bruyn \& Wilson(1987)]{deBruyn78}  de Bruyn, A.~G. \& Wilson,
A.~S. 1978, A\&A, {64}, 433

\bibitem[Dopita(2000)]{Do00}  Dopita M. A., 2000, \apss, 272, 79

\bibitem[Dopita et al.(1998)]{Do98}  Dopita, M.A., Heisler, C., Lumsden, S.,
\& Bailey, J., 1998, \apj,498, 570

\bibitem[Dopita et al.(2000)]{DoK00}  Dopita, M. A., Kewley, L. J., Heisler,
C. A., \& Sutherland, R. S., 2000, \apj, 542, 224

\bibitem[Dopita \& Sutherland(1995)]{do:su95}  Dopita, M. A., \& Sutherland,
R. S., 1995, \apj, 455, 468

\bibitem[Dopita \& Sutherland(1996)]{do:su96}  Dopita, M. A., \& Sutherland,
R. S., 1996, \apjs, 102, 161

\bibitem[Dopita \& Sutherland(2000)]{do:su00}  Dopita, M. A., \& Sutherland,
R. S., 2000, \apj, 539, 742

\bibitem[Elitzur \& Ferland(1986)]{EliFer86}  {Elitzur}, M., \& {Ferland},
G.~J., 1986, \apj, 305, 35

\bibitem[Ferguson et~al.(1997)]{Ferg97}  Ferguson, J.~W., Korista, K.~T.,
Baldwin, J.~A., \& Ferland, G.~J.\ 1997, \apj, 487, 122

\bibitem[Jones, Tielens, \& Hollenbach(1996)]{Jones96} Jones,
A.~P., Tielens, A.~G.~G.~M., \& Hollenbach, D.~J.\ 1996, \apj, 469, 740 

\bibitem[Komossa \& Schulz(1997)]{KomSch97}  Komossa, S.~\& Schulz, H.\
1997, \aap, 323, 31

\bibitem[Kraemer \& Crenshaw(2000)]{KC2000}  Kraemer, S. B., \& Crenshaw, D.
M., 2000, \apj, 532, 266

\bibitem[Krolik et al.(1981)]{kro:a}  Krolik, J. H., McKee, C., \& Tarter,
C. B., 1981, \apj, 249, 422

\bibitem[Laor \& Draine(1993)]{LaorDr93}  {Laor}, A. \& {Draine}, B.~T.,
1993 \apj, 402, 441

\bibitem[Mathis, Rumpl \& Nordsieck(1977)]{MRN77}  Mathis, J. S., Rumpl, W.,
\& Nordsieck, K. H, 1977, \apj, 217, 425

\bibitem[Moorwood et al.(1996)]{moor96}  Moorwood, A. F. M., Lutz, D.,
Oliva, E., Marconi, A., Netzer, H., Genzel, R., Sturm, E., \& de Graauw, T.,
1996, \aap, 315, L109

\bibitem[Netzer \& Laor(1993)]{NL93}  Netzer, H., \& Laor, A., 1993, \apjl,
404, L51

\bibitem[Oliva et al.(1994)]{oliv94}  Oliva, E., Salvati, M., Moorwood, A.
F. M., \& Marconi, A., 1994, \aap, 288, 457

\bibitem[Osterbrock(1989)]{Ost89}  Osterbrock, D. E., 1989, Astrophysics of
Gaseous Nebulae and Active Galactic Nuclei, (University Science Books)

\bibitem[Pedlar et al.(1989)]{Pedlar89}  Pedlar, A., Meaburn, J., Axon,
D.~J., Unger, S.~W., Whittle, D.~M., Meurs, E.~J.~A., Guerrine, N. \& Ward,
M.~J. 1989, MNRAS, {238}, 863

\bibitem[Tadhunter, Robinson \& Morganti(1989)]{Tad89}  Tadhunter, C.N.,
Robinson, A. \& Morganti, R. 1989, ESO Workshop on Extranuclear Activity in
Galaxies, eds. E. J. A. Meurs \& R.A. E. Fosbury, (ESO: Garching bie
M\"{u}nchen), p293

\bibitem[Tielens, McKee, Seab, \& Hollenbach(1994)]{Tieletal94} Tielens,
A.~G.~G.~M., McKee, C.~F., Seab, C.~G., \& Hollenbach, D.~J.\ 1994, \apj, 431,
321 

\bibitem[Veilleux(1991a)]{V91a}  Veilleux, S., 1991a, \apjs, 75, 357

\bibitem[Veilleux(1991b)]{V91b}  Veilleux, S., 1991b, \apjs, 75, 383

\bibitem[Veilleux(1991c)]{V91c}  Veilleux, S., 1991c, \apj, 369, 331

\bibitem[Veilleux \& Osterbrock(1987)]{VO87}  Veilleux, S., \& Osterbrock,
D. E., 1987, \apjs, 63, 295

\bibitem[V\'eron-Cetty \& V\'eron(2000)]{VV00}  V\'{e}ron-Cetty, M. P. \&
V\'{e}ron, P., 2000, \aapr, 10, 81

\bibitem[Wilson \& Raymond(1999)]{wilray99}  Wilson A. S., \& Raymond J. C.,
1999, \apjl, 513, 119

\bibitem[Wilson \& Willis(1980)]{Wilson80}  Wilson, A.~S. \& Willis, A.~G
1980, ApJ, {240}, 429
\end{thebibliography}
\end{document}